
%
%
\font\tenbf=cmbx10

\font\eightrm=cmr8
\font\eightit=cmti8

\font\germ=eufm10
\def\o{\otimes}
\def\goth#1{\hbox{\germ #1}}
\def\sectiontitle#1\par{\vskip0pt plus.1\vsize\penalty-250
 \vskip0pt plus-.1\vsize\bigskip\vskip\parskip
 \message{#1}\leftline{\tenbf#1}\nobreak\vglue 5pt}
\def\qed{\hbox{${\vcenter{\vbox{
    \hrule height 0.4pt\hbox{\vrule width 0.4pt height 6pt
    \kern5pt\vrule width 0.4pt}\hrule height 0.4pt}}}$}}
\hsize=5.0truein
\vsize=7.8truein
\parindent=15pt
\nopagenumbers
\baselineskip=13pt
\headline{\ifnum\pageno=1\hfil\else%
{\ifodd\pageno\rightheadline \else \leftheadline\fi}\fi}
\def\rightheadline{\hfil\eightit
CTM and Quantum algebra
\quad\eightrm\folio}
\def\leftheadline{\eightrm\folio\quad
\eightit
O. Foda and T. Miwa
\hfil}
\voffset=2\baselineskip
\centerline{\tenbf
CORNER TRANSFER MATRICES AND}
\centerline{\tenbf
QUANTUM AFFINE ALGEBRAS}
\vglue 24pt
\centerline{\eightrm
OMAR FODA$\dagger$ and TETSUJI MIWA${}^\ast$
}
\baselineskip=12pt
\centerline{\eightit
}
\baselineskip=10pt
\centerline{\eightit
${}\dagger$
Institute for Theoretical Physics,
University of Nijmegen,
}
\baselineskip=10pt
\centerline{\eightit
6525 ED NIJMEGEN, The Netherlands.
}
\baselineskip=10pt
\centerline{\eightit
${}^\ast$
Research Institute for Mathematical Sciences,
}
\baselineskip=10pt
\centerline{\eightit
Kyoto University, Kyoto 606, Japan
}
\vglue 20pt
\centerline{\eightrm ABSTRACT}
{\rightskip=1.5pc
\leftskip=1.5pc
\eightrm\parindent=1pc
Let
$\cal{H}$
be the corner-transfer-matrix Hamiltonian for the six-vertex model
in the anti-ferroelectric regime.
It acts on the infinite tensor product
$W=V\o V\o V\o\cdots$,  where
$V$
is the 2-dimensional irreducible representation of
the quantum affine Lie algebra
$U_q\bigl(\widehat{\goth{sl}}\,(2)\bigr)$.
We observe that $\cal{H}$ is the derivation of
$U_q\bigl(\widehat{\goth{sl}}\,(2)\bigr)$,
and conjecture
that the eigenvectors of
$\cal{H}$
form the level-1 vacuum representation of
$U_q\bigl(\widehat{\goth{sl}}\,(2)\bigr)$.
We report on checks in support of our
conjecture.
\vglue12pt}
\baselineskip=13pt
%
%

%
%
\def\Figure(#1|#2|#3)
{\midinsert
\vskip #2
\hsize 9cm
\raggedright
\noindent
{\bf Figure #1\quad} #3
\endinsert}
%
%
\def\Table #1. \size #2 \caption #3
{\midinsert
\vskip #2
\hsize 7cm
\raggedright
\noindent
{\bf Table #1.} #3
\endinsert}
%
%
\def\eq#1\endeq
{$$\eqalignno{#1}$$}

\def\leq#1\endeq
{$$\leqalignno{#1}$$}
%
%
\def\qbox#1{\quad\hbox{#1}\quad}
%
%

%
%
\def\nbox#1{\noalign{\hbox{#1}}}
%
%

%
%

%
%

%
%
\font\germ=eufm10
\def\goth#1{\hbox{\germ #1}}
%
%

%
%

%
%
\def\Definition#1.#2{\smallskip\noindent {\sl Definition #1.#2\quad}}
%
%
\def\subsec<#1|#2>{\medskip\noindent{#1}\hskip8pt{\sl {#2}}\par\smallskip}
\def\sub<#1>{\medskip\noindent\hskip8pt{\sl {#1}\/.}\quad}
%
%
\def\qed{\qquad$\Fsquare(.2cm,{})$}
%
%
\def\abstract#1\endabstract{
\bigskip
\itemitem{{}}
{\bf Abstract.}
\quad
#1
\bigskip
}
\def\m@th{\mathsurround=0pt}

\def\fsquare(#1,#2){
\hbox{\vrule$\hskip-0.4pt\vcenter to #1{\normalbaselines\m@th
\hrule\vfil\hbox to #1{\hfill$\scriptstyle #2$\hfill}\vfil\hrule}$\hskip-0.4pt
\vrule}}

\def\addsquare(#1,#2){\hbox{$
	\dimen1=#1 \advance\dimen1 by -0.8pt
	\vcenter to #1{\hrule height0.4pt depth0.0pt%
	\hbox to #1{%
	\vbox to \dimen1{\vss%
	\hbox to \dimen1{\hss$\scriptstyle~#2~$\hss}%
	\vss}%
	\vrule width0.4pt}%
	\hrule height0.4pt depth0.0pt}$}}

\def\Fsquare(#1,#2){
\hbox{\vrule$\hskip-0.4pt\vcenter to #1{\normalbaselines\m@th
\hrule\vfil\hbox to #1{\hfill$#2$\hfill}\vfil\hrule}$\hskip-0.4pt
\vrule}}

\def\Addsquare(#1,#2){\hbox{$
	\dimen1=#1 \advance\dimen1 by -0.8pt
	\vcenter to #1{\hrule height0.4pt depth0.0pt%
	\hbox to #1{%
	\vbox to \dimen1{\vss%
	\hbox to \dimen1{\hss$~#2~$\hss}%
	\vss}%
	\vrule width0.4pt}%
	\hrule height0.4pt depth0.0pt}$}}

\def\hfourbox(#1,#2,#3,#4){%

\fsquare(0.3cm,#1)\addsquare(0.3cm,#2)\addsquare(0.3cm,#3)\addsquare(0.3cm,#4)}

\def\Hfourbox(#1,#2,#3,#4){%

\Fsquare(0.4cm,#1)\Addsquare(0.4cm,#2)\Addsquare(0.4cm,#3)\Addsquare(0.4cm,#4)}

\def\hthreebox(#1,#2,#3){%
	\fsquare(0.3cm,#1)\addsquare(0.3cm,#2)\addsquare(0.3cm,#3)}

\def\htwobox(#1,#2){%
	\fsquare(0.3cm,#1)\addsquare(0.3cm,#2)}

\def\vfourbox(#1,#2,#3,#4){%
	\normalbaselines\m@th\offinterlineskip
	\vtop{\hbox{\fsquare(0.3cm,#1)}
	      \vskip-0.4pt
	      \hbox{\fsquare(0.3cm,#2)}
	      \vskip-0.4pt
	      \hbox{\fsquare(0.3cm,#3)}
	      \vskip-0.4pt
	      \hbox{\fsquare(0.3cm,#4)}}}

\def\Vfourbox(#1,#2,#3,#4){%
	\normalbaselines\m@th\offinterlineskip
	\vtop{\hbox{\Fsquare(0.4cm,#1)}
	      \vskip-0.4pt
	      \hbox{\Fsquare(0.4cm,#2)}
	      \vskip-0.4pt
	      \hbox{\Fsquare(0.4cm,#3)}
	      \vskip-0.4pt
	      \hbox{\Fsquare(0.4cm,#4)}}}

\def\vthreebox(#1,#2,#3){%
	\normalbaselines\m@th\offinterlineskip
	\vtop{\hbox{\fsquare(0.3cm,#1)}
	      \vskip-0.4pt
	      \hbox{\fsquare(0.3cm,#2)}
	      \vskip-0.4pt
	      \hbox{\fsquare(0.3cm,#3)}}}

\def\vtwobox(#1,#2){%
	\normalbaselines\m@th\offinterlineskip
	\vtop{\hbox{\fsquare(0.3cm,#1)}
	      \vskip-0.4pt
	      \hbox{\fsquare(0.3cm,#2)}}}

\def\Hthreebox(#1,#2,#3){%
	\Fsquare(0.4cm,#1)\Addsquare(0.4cm,#2)\Addsquare(0.4cm,#3)}

\def\Htwobox(#1,#2){%
	\Fsquare(0.4cm,#1)\Addsquare(0.4cm,#2)}

\def\Vthreebox(#1,#2,#3){%
	\normalbaselines\m@th\offinterlineskip
	\vtop{\hbox{\Fsquare(0.4cm,#1)}
	      \vskip-0.4pt
	      \hbox{\Fsquare(0.4cm,#2)}
	      \vskip-0.4pt
	      \hbox{\Fsquare(0.4cm,#3)}}}

\def\Vtwobox(#1,#2){%
	\normalbaselines\m@th\offinterlineskip
	\vtop{\hbox{\Fsquare(0.4cm,#1)}
	      \vskip-0.4pt
	      \hbox{\Fsquare(0.4cm,#2)}}}

\def\twoone(#1,#2,#3){%
	\normalbaselines\m@th\offinterlineskip
	\vtop{\hbox{\htwobox({#1},{#2})}
	      \vskip-0.4pt
	      \hbox{\fsquare(0.3cm,#3)}}}

\def\Twoone(#1,#2,#3){%
	\hbox{
	\normalbaselines\m@th\offinterlineskip
	\vtop{\hbox{\Htwobox({#1},{#2})}
	      \vskip-0.4pt
	      \hbox{\Fsquare(0.4cm,#3)}}}}

\def\threeone(#1,#2,#3,#4){%
	\normalbaselines\m@th\offinterlineskip
	\vtop{\hbox{\hthreebox({#1},{#2},{#3})}
	      \vskip-0.4pt
	      \hbox{\fsquare(0.3cm,#4)}}}

\def\Threeone(#1,#2,#3,#4){%
	\normalbaselines\m@th\offinterlineskip
	\vtop{\hbox{\Hthreebox({#1},{#2},{#3})}
	      \vskip-0.4pt
	      \hbox{\Fsquare(0.4cm,#4)}}}

\def\Threetwo(#1,#2,#3,#4,#5){%
	\normalbaselines\m@th\offinterlineskip
	\vtop{\hbox{\Hthreebox({#1},{#2},{#3})}
	      \vskip-0.4pt
	      \hbox{\Htwobox({#4},{#5})}}}

\def\threetwo(#1,#2,#3,#4,#5){%
	\normalbaselines\m@th\offinterlineskip
	\vtop{\hbox{\hthreebox({#1},{#2},{#3})}
	      \vskip-0.4pt
	      \hbox{\htwobox({#4},{#5})}}}

\def\twotwo(#1,#2,#3,#4){%
	\normalbaselines\m@th\offinterlineskip
	\vtop{\hbox{\htwobox({#1},{#2})}
	      \vskip-0.4pt
	      \hbox{\htwobox({#3},{#4})}}}

\def\Twotwo(#1,#2,#3,#4){%
	\normalbaselines\m@th\offinterlineskip
	\vtop{\hbox{\Htwobox({#1},{#2})}
	      \vskip-0.4pt
	      \hbox{\Htwobox({#3},{#4})}}}

\def\twooneone(#1,#2,#3,#4){%
	\normalbaselines\m@th\offinterlineskip
	\vtop{\hbox{\htwobox({#1},{#2})}
	      \vskip-0.4pt
	      \hbox{\fsquare(0.4cm,#3)}
	      \vskip-0.4pt
	      \hbox{\fsquare(0.4cm,#4)}}}

\def\Twooneone(#1,#2,#3,#4){%
	\normalbaselines\m@th\offinterlineskip
	\vtop{\hbox{\Htwobox({#1},{#2})}
	      \vskip-0.4pt
	      \hbox{\Fsquare(0.4cm,#3)}
	      \vskip-0.4pt
	      \hbox{\Fsquare(0.4cm,#4)}}}

\def\a{\fsquare(0.3cm){1}\addsquare(0.3cm)(2)\addsquare(0.3cm)(3)}

\def\b{\hbox{%
	\normalbaselines\m@th\offinterlineskip
	\vtop{\hbox{\fsquare(0.3cm){2}}\vskip-0.4pt\hbox{\fsquare(0.3cm){2}}}}}

\def\c{\hbox{\normalbaselines\m@th\offinterlineskip%
	\vtop{\hbox{\a}\vskip-0.4pt\hbox{\b}}}}


\dimen1=0.4cm\advance\dimen1 by -0.8pt
\def\ffsquare#1{%
	\fsquare(0.4cm,\hbox{#1})}

\def\naga{%
	\hbox{$\vcenter to 0.4cm{\normalbaselines\m@th
	\hrule\vfil\hbox to 1.2cm{\hfill$\cdots$\hfill}\vfil\hrule}$}}

\def\vnaga{\normalbaselines\m@th\baselineskip0pt\offinterlineskip%
	\vrule\vbox to 1.2cm{\vskip7pt\hbox to
\dimen1{$\hfil\vdots\hfil$}\vfil}\vrule}

\def\dvbox{\hbox{\normalbaselines\m@th\baselineskip0pt\offinterlineskip\vbox{%
	  \hbox{$\ffsquare 1$}\vskip-0.4pt\hbox{$\vnaga$}\vskip-0.4pt\hbox{$\ffsquare
N$}}}}
%
%
\def\sec(#1){Sect.\hskip2pt#1}
\font\germ=eufm10
\def\goth#1{\hbox{\germ #1}}
\def\U(#1){U_q(\goth{G}_#1)}

\def\Z{{\bf Z}}
\def\as(#1){U_q\bigl({\widehat{\goth{sl}}\hskip2pt(#1)}\bigr)}
\def\s(#1){U_q\bigl({\goth{sl}\hskip2pt(#1)}\bigr)}
\def\pn{\par\noindent}
\def\e{\varepsilon}

\def\b{B^{k,l}}
\def\L{\Lambda}

\def\g{\gamma}
\def\df{{{\rm def}\atop{\raise2pt\hbox{=}}}}

\def\title#1{\medskip\noindent{\it #1}\quad}

\def\al{\alpha}

\def\La{\Lambda}

\def\b{\tilde b}

\def\e{\tilde e}

\def\l{\langle}
\def\r{\rangle}
\def\Z{\hbox{{\bf Z}}}

\def\Q{\hbox{{\bf Q}}}

\font\germ=eufm10
\def\goth#1{\hbox{\germ #1}}
\def\g{\goth{g}}

\def\sqr#1#2#3{{\vcenter{\vskip-#3pt\vbox{\hrule height.#2pt
   \vskip-.3pt\hbox{\vrule width.#2pt height#1pt \kern#1pt
   \vrule width.#2pt}
   \vskip-.3pt\hrule height.#2pt}}}}

\def\uq{U_q(\g)}
\def\a0{\al_{i_{0}}}

\def\P{{\cal P}}

\def\rightup#1{\smash{\mathop{\hbox to 7mm{\rightarrowfill}}
\limits^{\lower 3pt #1}}}
\def\mathpalette#1#2{\mathchoice{#1\displaystyle{#2}}
   {#1\textstyle{#2}}{#1\scriptstyle{#2}}{#1\scriptscriptstyle{#2}}}
\def\c@ncel#1#2{\ooalign{$\hfil#1\mkern1mu/\hfil$\crcr$#1#2$}}
\def\notni{\mathrel{\mathpalette{\c@ncel}{\ni}}}
\def\a0{\al_{i_{0}}}

\def\P{{\cal P}}

\def\rightup#1{\smash{\mathop{\hbox to 7mm{\rightarrowfill}}\limits^{#1}}}
\def\mathpalette#1#2{\mathchoice{#1\displaystyle{#2}}
   {#1\textstyle{#2}}{#1\scriptstyle{#2}}{#1\scriptscriptstyle{#2}}}
\def\c@ncel#1#2{\ooalign{$\hfil#1\mkern1mu/\hfil$\crcr$#1#2$}}
\def\notni{\mathrel{\mathpalette{\c@ncel}{\ni}}}

\font\eightrm=cmr8
\def\uq{U_q(\goth{g})}
\def\Matrix#1#2#3#4#5{
\dimen6=#5
\dimen7=\dimen6
\dimen10=\dimen6
\advance \dimen10 by 6pt
\divide \dimen6 by 2
\dimen8=\dimen6
\dimen9=\dimen6
\advance \dimen8 by 0.2pt
\advance \dimen9 by -0.2pt
#1\kern-1mm
\raise1.8pt\hbox{$
{\mathop{\hbox to \dimen10{
\vrule width \dimen7 depth0pt height.4pt
\hskip-\dimen8
\vrule height \dimen6 depth \dimen6 width.4pt
\hskip \dimen9
\hfill}}
\limits^{\textstyle #2}_{\textstyle #3}}$}
\kern-.7mm #4
}

\def\bw(#1,#2,#3,#4){\Matrix{#1}{#2}{#3}{#4}{0.8cm}}

\def\uq{U_q\bigl(\widehat{\goth{sl}}\,(2)\bigr)}

\def\o{\otimes}
\def\L{\Lambda}
\def\H{{\cal{H}}}
\def\GT{>\hskip-3pt>}
\def\P{{\cal{P}}}
\def\e{\varepsilon}
\def\u{\phi}
\def\p{\hbox{$(+)$}}
\def\m{\hbox{$(-)$}}

\def\sm{\sum_{k=1}^\infty}
\def\l{\lambda}
\def\b{\emptyset}
\def\pu{{\overline p}_{k\,k+1}}
\def\pl{{\underline p}_{\,k\,k+1}}
\def\uelement{\mathop{\hbox{
\setbox1=\hbox{$\cup$}
\dimen1=\wd1
\dimen3=0.2pt
\divide \dimen1 by 2
\advance \dimen1 by \dimen3
\dimen5=\dimen1
\advance \dimen5 by -0.4pt
$\copy1\hskip-\dimen1 \vrule height\ht1 depth\dp1 width0.4pt\hskip \dimen5$
}}}

\newdimen\ex
\ex.2326ex
\newdimen\z
\z0pt

\def\varinjlim{\mathop{\vtop{\ialign{$##$\cr
\hfil{\fam\z lim}\hfil\cr\noalign{\nointerlineskip}%
{-}\mkern-6mu\cleaders\hbox{$\mkern-2mu{-}\mkern-2mu$}\hfill
\mkern-6mu{\to}\cr\noalign{\nointerlineskip\kern-\ex}\cr}}}}

\def\varprojlim{\mathop{\vtop{\ialign{$##$\cr
\hfil{\fam\z lim}\hfil\cr\noalign{\nointerlineskip}%
{\leftarrow}\mkern-6mu\cleaders\hbox{$\mkern-2mu{-}\mkern-2mu$}\hfill
\mkern-6mu{-}\cr\noalign{\nointerlineskip\kern-\ex}\cr}}}}

\def\downtiefill{$\braceld\leaders\vrule\hfill\bracerd$}

\def\overtie#1{\mathop{\vbox{\ialign{##\crcr
    \noalign{\kern3pt}\downtiefill\crcr
    \noalign{\kern3pt\nointerlineskip}
    $\hfil\displaystyle{#1}\hfil$\crcr}}}\limits}
\def\w{\iota}
\def\uq{U_q\bigl(\widehat{\goth{sl}}\,(2)\bigr)}

\def\o{\otimes}
\def\L{\Lambda}
\def\H{{\cal{H}}}
\def\GT{>\hskip-3pt>}
\def\P{{\cal{P}}}
\def\e{\varepsilon}
\def\u{\phi}
\def\p{\hbox{$(+)$}}
\def\m{\hbox{$(-)$}}


\beginsection 1.
Introduction

\subsec<1.1.|Motivation and results>
Baxter's corner transfer matrix (CTM) [B1] has been providing
an effective method
for computing 1-point functions in exactly-solvable lattice models:
the eight-vertex model [B1], the hard hexagon model and its
generalizations [B1], [ABF], [DJKMO1],
and the six-vertex model and its generalizations [DJKMO2], [(KMN)${}^2$].

These computations revealed a deep connection between lattice
models and quantum affine Lie algebras, where the temperture variable
in the former is the deformation parameter
$q$ in the latter. Let us recall two examples of this connection.

The multiplicities of the CTM eigenvalues of the
six-vertex model and its generalizations are
equal to the weight multiplicities of irreducible highest weight
representations of a certain affine Lie algebra [DJKMO2].

The CTM eigenvectors, in the zero temperaturte limit,
coincide, as has been shown for a large class of vertex models,
with the base vectors (the crystal base [K])
of irreducible highest weight representations of the
quantum affine Lie algebras
$U_q(\g)$
at
$q=0$, [MM], [JMMO], [(KMN)${}^2$].

In this paper, we search for an algebraic structure in the
CTM eigenvectors at $q\neq0$.
We work in the context of a simple off-critical model:
the six-vertex model in the anti-ferroelectric
regime (cf. 8.10, [B1]). For this model
the corresponding algebra is
$\uq$
and the corresponding representation is the level-1 vacuum representation.

For clarity, we present our results basically in the order that
we obtained them. We first observed that, though the \lq bare\rq\
CTM Hamiltonian, in the large lattice limit, has divergent eigenvalues,
one can define a
\lq renormalized\rq\  Hamiltonian, order by order in a low-temperature
expansion, that has finite eigenvalues bounded from below.
We report on that in \S 2.

Because of the identification, mentioned above,
of the eigenvectors at
$q=0$
and the crystal base,
it is natural to suspect that, even at
$q\neq0$,
the eigenvectors of $\H$ are identified with
the weight vectors in the level-1 vacuum representation of
$\uq$.

Our preliminary checks confirmed that
the ground state (the lowest eigenvector of $\H$)
satisfies the highest weight conditions of $\uq$,
as a vector in the semi-infinite tensor product of the two-dimensional
representation of $\uq$, on which $\H$ acts naturally.
So we proceeded to compute the highest weight vector and
the next few descendants in this tensor representation.
This computation is independent of that
used to compute the eigenvectors of $\H$. We report on that in \S 3.

Once we had these two, {\it a priori} independent sets of vectors,
we proceeded systematically to compare them. This is discussed in \S 4.
Our computations confirm that the eigenvectors of $\H$ form indeed
the level-1 representation of $\uq$. Though non-trivial, and
quite extensive, these computations are
limited to a finite number of terms in a perturbation expansion,
and we can only make the above statement as a conjecture.

For further support of our conjecture, we observe that
$\H$
satisfies simple commutation relations with the generators
of the quantum affine Lie algebra $\uq$. More precisely:
$\H$ acts as a derivation of $\uq$.

In the following subsection we wish to recall
the $q=0$ result.

\subsec<1.2.|The low temperature limit $q=0$>
The six-vertex model, that we are interested in, is related
to the quantum affine Lie algebra $\uq$ of Drinfeld,
and Jimbo, [D] [J], since its Boltzmann weights form
the $R$-matrix of the 2-dimensional representation $V$,
of $\uq$. As we mentioned above,
the deformation parameter $q$ has the physical
interpretation of temperature. We restrict our consideration
to the anti-ferroelectric regime:
$0<x<1,-1<q<0$, i.e., in particular,
$\Delta=(q+q^{-1})/2<-1$.
This is an off-critical, ordered regime.
In the limit $q \rightarrow 0_{-}$, $\Delta \rightarrow - \infty$,
it is totally ordered, and things become simpler.

Kashiwara's theory of the crystal bases
of quantum groups [K]: the representation theory of
$\uq$
at
$q=0$,
enables us to sum the
$q$-series
obtained by the CTM method,
and to compute the 1-point functions [(KMN)${}^2$].
We wish to elaborate this point.

Let
$A(x,q)$
be a CTM for the six-vertex model.
The parameter
$x$
is the multiplicative spectral parameter.
In the large lattice limit, the CTM has the asymptotic form [B1]
\eq
&A(x,q)=\alpha(x,q)x^{\H(q)}.&(1.2.1)\cr
\endeq
Here
$\alpha(x,q)$
is a normalization constant and
$\H=\H(q)$
is the CTM Hamiltonian.
It acts naturally on the infinite tensor product
\eq
&W=V\o V\o V\o\cdots.&(1.2.2)\cr
\endeq
{}From the periodicity requirement for the eigenvalues of the CTM
one can deduce that the eigenvalues of
$\H$
are
$0,1,2,\cdots$.
The multiplicities of these eigenvalues are
equal to the weight multiplicities of the level-1 vacuum
representation
$V(\L_0)$ [DJKMO2].
Since the multipicity is independent of
$q$,
we can compute it in the limit of
$q=0$.
In fact, at
$q=0$,
the CTM Hamiltonian
$\H$ is diagonal with respect to pure tensors in
$W$.

A pure tensor, or more appropriately,
a {\sl path}, as we will refer to it
from now on, is a vector
$|p\r$
in
$W$
of the form
\eq
&|p\r=p(1)\o p(2)\o p(3)\o\cdots,&(1.2.3)\cr
\endeq
where
$p(k)=(\pm)$
is the weight vector
$\p$
or
$\m$
in the
$k$-th component
$V=\Q(q)\p\oplus\Q(q)\m$
of
$W$.
Think of $W$ as a semi-infinite \lq spin-chain\rq: there are
two possible, equivalent anti-ferroelectric ground states.
We choose to work with the one that satisfies the following
boundary conditions
\eq
p(k)&=\p\qbox{if $k$ is even and $\GT0$.}&(1.2.4)\cr
&=\m\qbox{if $k$ is odd and $\GT0$.}\cr
\endeq
Let us denote by
$\P$
the set of all paths that satisfy (1.2.4). These
paths are the eigenvectors of the CTM in the
low-temperature limit
$q\rightarrow{0}_{-}$.

In [K] the crystal base
$B(M)$
was introduced for an integrable representation $M$ of
$U_q(\goth{g})$.
It is a base of
$M$
at
$q=0$
that enjoys remarkably simple combinatorial properties.

Consider the level-1 vacuum representation
$M=V(\La_0)$ for $\uq$.
There is a natural isomorphism between the set
$\P$
and the crystal
$B\bigl(V(\L_0)\bigr)$ [MM] (see also [JMMO] and [(KMN)${}^2$]).
In other words, the paths, which are the eigenvectors of the CTM at
$q=0$,
are at the same time the base of
$V(\La_0)\subset W$
at
$q=0$.

In this work, we take a first step to extend the above result to
$q\neq0$.
We infer, on the basis of direct computations, that
the subspace of
$W$
spanned by the eigenvectors
\eq
&|v\r=\sum_p c(p,v)|p\r&\cr
\endeq
of the CTM is isomorphic to
$V(\La_0)$
under the action of
$\uq$ ---the action {\it via} the infinite comultiplication on
$W$.

\subsec<1.3.|Related work>

In [B2], Baxter discussed the eight-vertex model.
In particular, at the point where it reduces to two decoupled Ising models,
the diagonalization of the CTM by means of the spinor representation is given.
There are more results in [Da1], [TP], [IT] for the Ising case,
also on the basis of the free fermion structure of the Ising model.

As for the six-vertex model,
in [Da2], the diagonalization of the CTM Hamiltonian is studied in the
ferroelectric phase.

Interesting observations are given in [Te]: the CTM Hamiltonian
is a boost operator for the row-to-row transfer matrix (cf. [SW], [T]),
and in [SB]: there is a similarity in the spectrum of the CTM
in the thermodynamic limit at an off critical temperature
and that of the row-to-row transfer matrix on a finite size lattice
at the critical temperature.
The latter was further discussed in [DP], [F].

\subsec<{}|Acknowledgements>
The first author wishes to thank the organizers of
\lq RIMS91 project on INFINITE ANALYSIS\rq\  for their
friendly hospitality, and for the pleasant atmosphere
at RIMS. We both wish to thank the participants
of the project for discussions, and Brian Davies for useful
communications. We are indebted to Atsushi Nakayashiki
for careful reading of the manuscript.
\def\bmod{\hbox{\rm mod\ }}
\def\eg{\omega}
\def\sm{\sum_{k=1}^\infty}
\def\l{\lambda}
\def\b{\hbox{\rm\O}}
\def\pu{Q^+_kp}
\def\pl{Q^-_kp}
\def\sg#1{\hbox{$(#1)$}}
\def\lunit#1{\bigl(#1\o1\bigr)}
\def\runit#1{\bigl(1\o#1\bigr)}

\beginsection 2.
The eigenvectors of $\H$

\subsec<2.1.|The model>
We consider the six-vertex model in the anti-ferroelectric regime,
which is one of the simpler off-critical lattice models with an exact
solution [B1]. The variables of the model are arrows that live on the
bonds of a lattice, and can point in either direction.
In this paper we use \lq signs\rq\ $(\pm)$ instead of
\lq arrows\rq\  as we shall explain below.

Draw the lattice diagonally, i.e., with bonds to be directed SW-NE or SE-NW.
Let $\p$ (resp. $\m$) stand for an arrow that points upwards (resp. downwards).
Figure 2.1 shows an equivalent pair of signs and arrows around a vertex.
To each configuration of signs around a vertex we associate a Boltzmann weight.
One can think of the signs in the upper row as an initial configuration that
develops to a final configuration given by the signs in the lower row,
with a probability given by the corresponding Boltzmann weight.

\Figure(2.1.1|4cm|A configuration around a vertex.)

The non-vanishing Boltzmann weights,
written as rotated by $45^\circ$ clockwise, are
\eq
&\bw(\p,\p,\p,\p)=\bw(\m,\m,\m,\m)={x-q^2\over1-q^2},&\cr
&\bw(\m,\p,\p,\m)=\bw(\p,\m,\m,\p)={q(x-1)\over1-q^2},\cr
&\bw(\p,\p,\m,\m)=x,\bw(\m,\m,\p,\p)=1.&(2.1.1)\cr
\endeq
All other possible configuration have vanishing Boltzmann weights.

The parametrization (2.1.1) has the following meaning
in connection with the $R$-matrix of $\uq$.
Let
$V$
be the 2-dimensional
$\Q(q)$
vector space with the distinguished base vectors
$(+)$
and
$(-)$.
Introduce a matrix
$R(x,q)$
by
\eq
&R(x,q)(i)\o(j)=\sum_{k,l}\bw(\sg k,\sg i,\sg l,\sg j)(k)\o(l).&(2.1.2)\cr
\endeq
Then it satisfies the Yang-Baxter equation:
\eq
&\runit{R(x,q)}\lunit{R(xy,q)}\runit{R(y,q)}\cr
&=\lunit{R(y,q)}\runit{R(xy,q)}\lunit{R(x,q)}.\cr
\endeq

\subsec<2.2.|Definitions>
Let
$A(x,q)$
be the CTM of the SE quadrant [B1].
In the large lattice limit,
it is of the form (1.2.1),
where
$\H=\H(q)$
acts on
$W$
given by (1.2.2)
with
$V=\Q(q)(+)\oplus\Q(q)(-)$.

Let us compute
$\H$.
Set
$x=e^\l$.
Note that the
$R$-matrix
given by (2.1.1-2) is proportional to the identity matrix at
$\l=0$.
Let
$R(x,q)=1+r(q)\l+O(\l^2)$
be the expansion of
$R(x,q)$
at
$\l=0$.
We have
\eq
&\H=\sm kr(q)_{k,k+1}.&\cr
\endeq
Here
$r(q)_{k,k+1}$
means
$r(q)$
acting on the
$k$-th
and the
$(k+1)$-th
components of
$W$.
We expect that the difference of any two eigenvalues of
$\H$
is an integer.

Let us compute
$\H$
explicitly.
We denote by
$(s\ s')$
the $k$-th and the $(k+1)$-th components of a vector in
$W$.
Define two operators
$P'_k$
and
$Q_k$
that act on $(s\ s')$. The prime
in $P'_k$ is because this operator will be re-defined
shortly.

The diagonal operator
$P'_k$
is given by
\eq
P'_k(s\ s')&=H(s,s')(s\ s')&(2.2.1)\cr
\nbox{where}
H(s,s')&=1\hskip9pt\qbox{if $(s\ s')=(+\ +),(+\ -),(-\ -)$,}\cr
&=0\hskip10pt\qbox{if $(s\ s')=(-\ +)$.}\cr
\endeq
The operator
$Q_k$
is given by
\eq
&Q_k(\pm\ \mp)=(\mp\ \pm),\quad Q_k(\pm\ \pm)=0.&(2.2.2)\cr
\endeq
In other words, it permutes adjacent spins that are different.

We also define an operator
$S_k$ that measures spin. It acts
on the
$k$-th component of
$W$ as
\eq
&S_k\p={1\over2}\p,\ S_k\m=-{1\over2}\m.&\cr
\endeq
{}From the Boltzmann weights, and the above definitions,
one can show that the Hamiltonian
$\H$ is given by
\eq
&\H={1\over 1-q^2}\left\{\sm k\bigl\{P'_k+q^2(P'_k-1)\bigr\}
+q\sm kQ_k-q^2\sm S_k\right\}.&\cr
\endeq

We call the path
\eq
&|\b\r=(-\ +\ -\ +\cdots)&\cr
\endeq
the ground-state path: or the zero-temperature ground state.
The point is that, as we will see,
at $q \neq 0$, the true ground state (i.e., the lowest eigenvector of
$\H$)
consists of an infinite linear
superposition of paths, with $q$-dependent coefficients. At $q = 0$
the ground state reduces to the above path
$|\b\r$.
The coefficients of all others vanish.

We fix the value of the spin operator
$S=\sm S_k$
on
$|\b\r$
as
$S|\b\r=0$.
Consequently,
$S$
is well-defined on all the paths in $\P$.
We call the eigenvalue of
$S$
at
$p$
the spin
of
$p$
and denote it by
$s(p)$.
We define
\eq
&\w(p)=\min\{n\mid\exists i_1,\ldots,i_n\hbox{ such that }
|p\r=Q_{i_1}\cdots Q_{i_n}|\b\r\},&\cr
\nbox{and}
&l(p)=\max\{n\mid p(n)\neq\b(n)\}.&\cr
\endeq
We refer to  them as the depth and the length of
$p$, respectively.
The depth is the \lq inversion number\rq\  of
$p$
relative to the ground state path: that is, starting from
the ground state path, it counts the minimal number of inversions
one has to perform on neighbouring pairs $(+,-)$ or $(-,+)$ to obtain
a certain path.

The length gives the extension of spin fluctuations in
$p$, as compared to the ground state path: that is,
it gives the left-most spin position following which
a given path coincides with the ground state path.

\subsec<2.3.|The renormalized Hamiltonian>
Starting to compute the eigenvectors of
the CTM Hamiltonian given above, one finds
immediately that its eigenvalues
are infinite. That is why we refer to $\H$ as the
\lq bare\rq\  Hamiltonian. We need to renormalize it.

First, we look for the lowest eigenvector
$|\u\r$. In all generality, we expect it to be
of the form
\eq
&|\u\r=\sum_pc(p,\u)|p\r,&(2.3.1)\cr
\endeq
where the sum is restricted to such paths that
$s(p)=0$.

We are going to subtract an infinite scalar from
$\H$
in such a way that the lowest eigenvalue is shifted to $0$.
We will surmise that this infinite scalar can be
defined consistently, order-by-order
in a pertubation expansion in $q$, and refer to the final result
as the \lq renormalized\rq\  Hamiltonian.

The first step is to redefine $P'_k$ so that the energy
of the ground state path is $0$. We define
$P_k$
by
\eq
P_k(s\ s')&=\bigl(H(s,s')-H(-,+)\bigr)(s\ s')\qbox{if $k$ is odd,}&\cr
&=\bigl(H(s,s')-H(+,-)\bigr)(s\ s')\qbox{if $k$ is even.}\cr
\endeq
We define the \lq classical\rq\  energy
$\eg(p)$ of a path $p$
to be the eigenvalue of
$P=\sm kP_k$
at
$p$. This is the actual energy of a path
at $q = 0$.

We suppose the renormalized Hamiltonian is of the form
\eq
&\H_{re}={1+q^2\over1-q^2}(P+\e Q-\e qS-\e R),&(2.3.2)\cr
\nbox{where}
&Q=\sm kQ_k,\ R=\sm kR_k,\ \e={1\over2\Delta}.\cr
\endeq
Here
$R_k$
is a scalar to be determined in the recursive process of computing
$c(p,\u)$.
We wish to choose
$R$
so that the lowest eigenvalue of
$\H_{re}$
is zero.

\subsec<2.4.|Spin-0 paths>
Denoting the paths in terms of $+$ and $-$ spins is not
convenient for long paths.
Let us introduce a more concise notation.
Given a path $p$ of spin-0,
we define a list of non-negative integers
\eq
&a=[a(1),a(2),\cdots]\cr
\endeq
as follows.
Let us introduce the  auxiliary parameters
$k$,
$s$
and
$c$.
We start from
$k=0,a=[\ ],s=-1,c=0$.
We increase
$k$.
At the
$k$-th process we do the following;
{\obeylines\medskip
if $s=p(k)$ then
\quad if $c=0$ then
\quad\quad (i) add $0$ to $a$ at the last column
\quad\quad(ii) change $s$ to $-s$
\quad else
\quad\quad (i) change $c$ to $c-1$
else\hskip4.1in(2.4.1)
\quad (i) add $c+1$ to $a$ at the last column
\quad(ii) change $c$ to $c+1$}
\medskip
\pn
The list
$a$
satisfies
\eq
&a(i)\ge0,a(i+1)\le a(i)+1,a(i)=0\hbox{ if }i\GT0.&(2.4.2)\cr
\endeq
The correspondence between the spin-0 paths
and the lists satisfying (2.4.2) is one-to-one.
We label
$p$
as
$(a(i))_{1\le i\le m}$
where
$m=\max\{i\mid a(i)\neq0\}$.
The following are the paths of spin 0 and length less than 6.

\settabs\+\indent
&\hbox{$|00001\r$}\quad
&\hbox{$-\ +\ -\ +\ -\ +\ -\ +\ \cdots$}\quad
&\quad1\quad
&\cr
\+&\hbox{$a$}\hfill
&\hfill\hbox{$p$}\hfill
&\hbox{$\w(p)$}\hfill
&\hbox{$\eg(p)$}\cr
\noindent$l=0$
\+&\hbox{$|\b\r$}        &\hbox{$-\ +\ -\ +\ -\ +\ -\ +\ \cdots$}&0&0\cr
\noindent$l=2$
\+&\hbox{$|1\r$}         &\hbox{$+\ -\ -\ +\ -\ +\ -\ +\ \cdots$}&1&1\cr
\noindent$l=3$
\+&\hbox{$|01\r$}        &\hbox{$-\ -\ +\ +\ -\ +\ -\ +\ \cdots$}&1&2\cr
\noindent$l=4$
\+&\hbox{$|001\r$}       &\hbox{$-\ +\ +\ -\ -\ +\ -\ +\ \cdots$}&1&3\cr
\+&\hbox{$|11\r$}        &\hbox{$+\ -\ +\ -\ -\ +\ -\ +\ \cdots$}&2&2\cr
\+&\hbox{$|12\r$}        &\hbox{$+\ +\ -\ -\ -\ +\ -\ +\ \cdots$}&3&4\cr
\noindent$l=5$
\+&\hbox{$|0001\r$}      &\hbox{$-\ +\ -\ -\ +\ +\ -\ +\ \cdots$}&1&4\cr
\+&\hbox{$|011\r$}       &\hbox{$-\ -\ +\ -\ +\ +\ -\ +\ \cdots$}&2&3\cr
\+&\hbox{$|012\r$}       &\hbox{$-\ -\ -\ +\ +\ +\ -\ +\ \cdots$}&3&6\cr
\+&\hbox{$|101\r$}       &\hbox{$+\ -\ -\ -\ +\ +\ -\ +\ \cdots$}&2&4\cr
\noindent$l=6$
\+&\hbox{$|00001\r$}     &\hbox{$-\ +\ -\ +\ +\ -\ -\ +\ \cdots$}&1&5\cr
\+&\hbox{$|0011\r$}      &\hbox{$-\ +\ +\ -\ +\ -\ -\ +\ \cdots$}&2&4\cr
\+&\hbox{$|0012\r$}      &\hbox{$-\ +\ +\ +\ -\ -\ -\ +\ \cdots$}&3&8\cr
\+&\hbox{$|0101\r$}      &\hbox{$-\ -\ +\ +\ +\ -\ -\ +\ \cdots$}&2&7\cr
\+&\hbox{$|1001\r$}      &\hbox{$+\ -\ -\ +\ +\ -\ -\ +\ \cdots$}&2&6\cr
\+&\hbox{$|111\r$}       &\hbox{$+\ -\ +\ -\ +\ -\ -\ +\ \cdots$}&3&3\cr
\+&\hbox{$|112\r$}       &\hbox{$+\ -\ +\ +\ -\ -\ -\ +\ \cdots$}&4&7\cr
\+&\hbox{$|121\r$}       &\hbox{$+\ +\ -\ -\ +\ -\ -\ +\ \cdots$}&4&5\cr
\+&\hbox{$|122\r$}       &\hbox{$+\ +\ -\ +\ -\ -\ -\ +\ \cdots$}&5&6\cr
\+&\hbox{$|123\r$}       &\hbox{$+\ +\ +\ -\ -\ -\ -\ +\ \cdots$}&6&9\cr

\subsec<2.5.|A perturbative expansion>
We are going to solve
\eq
&\H_{re}|\u\r=0&(2.5.1)\cr
\endeq
perturbatively at
$\e=0$.
($\H_{re}$
is given in (2.3.2).)
We assume the following functional form of the coefficients in (2.3.1):
\eq
&c(\b,\u)=1,&\cr
&c(p,\u)=0\hbox{ if }s(p)\neq0,\cr
&c(p,\u)=\sum_{l\ge0}c_l(p,\u)\e^{2l+\w(p)}.&(2.5.2)\cr
\endeq
We set
$c_l(0,\u)=0$
for later convenience.
We assume also that
\eq
&R_k=\sum_{j=0}^\infty r_{jk}\e^{2j+1}&\cr
\endeq
where the coefficients
$r_{jk}$
are yet unknown.

We define the operators $Q^+_k$ and $Q^-_k$ as follows.
\eq
Q^\pm_k|p\r&=Q_k|p\r\hbox{ if }Q_k|p\r\neq0\hbox{ and }\w(Q_kp)=\w(p)\pm1,&\cr
&=0\hskip20pt\hbox{ otherwise.}\cr
\endeq

\sub<Example>
If $|p\r=|011\r$, the expression
$c_0(\pu,\u)+c_1(\pl,\u)$
means
$0$, $c_1(0001,\u)$, $c_0(012,\u)$, $c_1(01,\u)$, $0$, $\ldots$
when
$k=1,2,3,4,5,\ldots$, respectively.
\medskip
With this notations, (2.5.1) reads as
\eq
\eg(p)c_l(p,\u)&+\sum_{k=1}^\infty
k\bigl\{c_{l-1}(\pu,\u)+c_l(\pl,\u)\bigr\}&\cr
&-\sum_{k=1}^\infty\sum_{j=0}^\infty kr_{jk}c_{l-1-j}(p,\u)=0.&(2.5.3)\cr
\endeq

Let us determine the unknown terms
$r_{jk}$.
Consider (2.5.3) for
$|p\r=|\b\r$.
Then, by using (2.5.2) we have
\eq
&\sum_kkc_{l-1}(\underbrace{0\cdots0}_{k-1}1,\u)
=\sum_kkr_{l-1\,k}&\cr
\endeq
Therefore, we choose
\eq
&R_k=c(\underbrace{0\cdots0}_{k-1}1,\u),&(2.5.4)\cr
\nbox{or equivalently}
&r_{jk}=c_j(\underbrace{0\cdots0}_{k-1}1,\u).\cr
\endeq
With this choice the recursion equation (2.5.3)
becomes self-contained.
It can be solved self-consistently if and only if
the succesive coefficients satisfy the asymptotic factorizability: for any
$j$
the equality
\eq
&c(\pu,\u)\equiv
c(\underbrace{0\cdots0}_{k-1}1,\u)c(p,\u)\bmod\e^j&(2.5.5)\cr
\endeq
holds for sufficiently large
$k$.
In fact, if (2.5.5) is valid
(and that is certainly the case according to our extensive
computer checks),
we can truncate the summation on
$k$
in (2.5.3) at a sufficiently large value of $k$,
and compute
$c_l(p,\u)$
independently of the cut-off.

\sub<Example>
Let us illustrate how we perturbatively solve (2.5.3).
Take
$|p\r=|1\r$.
Then, for $l=0$, (2.5.3) reads as
$c_0(1,\u)+c_0(\b,\u)=0$.
Therefore, we have
$c_0(1,\u)=-1$.
Similarly, we have
$r_{0k}=-1$
for all
$k$.
Therefore, for $l=1$, (2.5.3) reads as
\eq
c_1(1,\u)&+c_1(\b,\u)+3c_0(11,\u)+4c_0(101,\u)+5c_0(1001,\u)+\cdots\cr
&+(1+2+3+4+5+\cdots)c_0(1,\u)=0.\cr
\endeq
Using that
$c_1(\b,\u)=0$, $c_0(1,\u)=-1$,
$c_0(11,\u)=2$, $c_0(101,\u)=1$, $c_0(1001,\u)=1,\ldots\,$,
which are computable prior to
$c_1(1,\u)$,
we have
$c_1(1,\u)=0$.

\subsec<2.6.|Excited states>
Take
$s\in\Z$ and $\eg\in\Z_{>0}$.
An excited state with spin
$s$
and energy
$\eg$
is a formal sum
$|v\r=\sum_pc(p)|p\r$
satisfying
\eq
&S|v\r=s|v\r,&(2.6.1)\cr
&\H_{re}|v\r=\eg|v\r.&(2.6.2)\cr
\endeq
Let us denote the space of such vectors by
$H_{s,\eg}$.

Suppose that there are
$m$
distinct paths
$p_1$, $p_2,\ldots,p_m$
that have the same values of the spin and the classical energy.
Then, one can expect that there are
$m$
distinct solutions to the eigenvalue equations (2.6.1-2).
We use the following Ansatz:
choose $i\in\{1,2,\ldots,m\}$, and impose that
\eq
&c(p_j)=\delta_{ij},&\cr
&c(p)=\sum_{l\ge0}c_l(p)\e^{\w(p,p_i)+2l}.&(2.6.3)\cr
\endeq
Here we denote by
$\w(p,p')$
the inversion number of
$p$
relative to
$p'$.
For example,
$\w(-+++-+\cdots,++-+-+\cdots)=2$.

The recursion equation reads as
\eq
\bigl(\eg(p)&-\eg\bigr)c_l(p)
+\sum_k\bigl(c_{l-1}(\pu)+c_l(\pl)\bigr)&\cr
&-\sum_k\sum_{j=0}^{l-1}kr_{jk}c_{l-1-j}(p)
=\sum_{j=1}^l\bigl(\eg a_j+sb_{j-1}\bigr)c_{l-j}(p).&(2.6.4)\cr
\endeq
Here
\eq
&\sqrt{1-4\e^2}=\sum_{j=0}^{\infty}a_j\e^{2j}\left(={1-q^2\over1+q^2}\right),\
{1-\sqrt{1-4\e^2}\over2\e}=\sum_{j=0}^{\infty}b_j\e^{1+2j}(=q).\cr
\endeq
We denote the eigenvector obtained from (2.6.3-4) by
$|v(p_i)\r$.

We have
$\eg(p)-\eg=0$
for
$p=p_j$.
Therefore, for each
$j\in\{1,2,\ldots,m\}$
and each
$l$,
we get
a consistency condition from (2.6.4).
We have checked extensively
that these consistency conditions are indeed satisfied.
An analytic proof is outside the limited scope of this paper.
\def\s{\sigma}
\def\D{\Delta}
\def\I{\Delta^{(\infty)}}
\def\P{{\cal{P}}}
\def\E{\e}
\def\Xa{1}
\def\Xb{-\E}
\def\Xc{\E^3-\E}
\def\Xf{\E^5-\E^3}
\def\Xe{\E^6-2\E^4+2\E^2}
\def\Xd{-\E^5+\E^3-\E}
\def\Xi{\E^9-3\E^7+3\E^5-\E^3}
\def\Xh{\E^{10}-3\E^8+5\E^6-5\E^4+2\E^2}
\def\Xg{\E^9-\E^7+\E^3-\E}
\def\Xj{-2\E^8+3\E^6-2\E^4+\E^2}
\def\Xt{-\E^{12}+3\E^{10}-3\E^8+\E^6}
\def\Xs{-2\E^{13}+7\E^{11}-11\E^9+9\E^7-3\E^5}
\def\Xr{-2\E^{14}+7\E^{12}-11\E^{10}+11\E^8-8\E^6+3\E^4}
\def\Xq{-\E^{14}+4\E^{12}-10\E^{10}+13\E^8-9\E^6+3\E^4}
\def\Xp{-\E^{15}+4\E^{13}-10\E^{11}+17\E^9-19\E^7+13\E^5-5\E^3}
\def\Xo{-\E^{14}+5\E^{12}-8\E^{10}+7\E^8-3\E^6+\E^2}
\def\Xm{\E^{13}-2\E^{11}+4\E^9-5\E^7+3\E^5-\E^3}
\def\Xl{\E^{14}-2\E^{12}+3\E^{10}-6\E^8+8\E^6-5\E^4+2\E^2}
\def\Xk{\E^{13}-5\E^{11}+7\E^9-4\E^7+\E^3-\E}
\def\Xn{-2\E^{12}+6\E^{10}-8\E^8+6\E^6-3\E^4+\E^2}
\def\v{\hbox{\rm vac}}


\beginsection 3.
The highest weight vector and descendants in $W$

\subsec<3.1.|Formal $\uq$ action>
To recall, the object of our interest is the semi
infinite tensor product $W$ (1.2.2),
of copies of the 2-dimensional irreducible
representation $V$, of $\uq$.
We can think of the weight vectors
$(+)$ and $(-)$
of $V$ as spin-up and spin-down,
or equivalently, as
spin-$\displaystyle{1\over2}$
and
spin-$\displaystyle{-1\over2}$.

The action of the generators of
$\uq$
on
$V$
is as follows:
\eq
&f_0(+)=e_1(+)=0,\ f_0(-)=e_1(-)=(+),&\cr
&f_1(+)=e_0(+)=(-),\ f_1(-)=e_0(-)=0,\cr
&t_1(+)=t_0^{-1}(+)=q(+),\ t_1(-)=t_0^{-1}(-)=q^{-1}(-).\cr
\endeq
Notice that the generators form pairs that have
the same action on the spin variables.

In our conventions, co-multiplication is of the form:
\eq
&\D(e_i)=e_i\o1+t_i^{-1}\o e_i,&\cr
&\D(f_i)=f_i\o t_i+1\o f_i,\cr
&\D(t_i)=t_i\o t_i.\cr
\endeq
Formally, the action of
$e_i$, $f_i$ and $t_i$
on
$W$
is as follows:
\eq
&\I(e_i)=e_i\o1\o1\o\cdots+t_i^{-1}\o e_i\o1\o\cdots+\cdots,\cr
&\I(f_i)=f_i\o t_i\o t_i\o\cdots+1\o f_i\o t_i\o\cdots+\cdots,\cr
&\I(t_i)=t_i\o t_i\o t_i\o\cdots.&(3.1.1)\cr
\endeq
{}From now on we abbreviate
$\I(e_i)$, $\I(f_i)$ and $\I(t_i)$
to
$e_i$, $f_i$ and $t_i$,
respectively.

We have
$t_0t_1=1$
on
$V$.
Since we are aiming at constructing the level-1 vacuum representation in
$W$,
we define
\eq
&t_0|p\r=q^{1-2s(p)}|p\r,\quad t_1|p\r=q^{2s(p)}|p\r.&\cr
\endeq
Therefore,
$t_0t_1=q$
on
$W$.

Consider a formal element of
$W$:
\eq
&|v\r=\sum_pc(p,v)|p\r.&(3.1.2)\cr
\endeq
Our point is that, at $q \neq 0$, the physical states are infinite linear
superpositions of paths in $W$.
We use the convension
$c(0,v)=0$.

Let us introduce a partial net spin
$s_k(p)$
by
\eq
&s_k(p)=\sum_{1\le i\le k-1}{1\over2}p(i).&\cr
\endeq
We define a map
$\s^\pm_k:\P\rightarrow\P\sqcup\{0\}$
by
\eq
\s^\pm_k(p)&=0\quad\hskip12pt\hbox{if }p(k)=\pm,&\cr
\s^\pm_k(p)(j)&=p(j)\quad\hbox{if }j\neq k\hbox{ and }p(k)=\mp,&\cr
&=(\pm)\quad\hskip2pt\hbox{if }j=k\hbox{ and }p(k)=\mp.&\cr
\endeq
Note that
$s\bigl(\s^\pm_k(p)\bigr)=s(p)\pm1$.

Formally, the actions of
$e_i$
and
$f_i$
given by (3.1.1) can be expressed as follows:
\eq
&c(p,e_0v)=\sum_kq^{2s_k(p)}c\bigl(\s^+_k(p),v\bigr),&\cr
&c(p,e_1v)=\sum_kq^{-2s_k(p)}c\bigl(\s^-_k(p),v\bigr),\cr
&c(p,f_0v)=\sum_kq^{2-2s(p)+2s_k(p)}c\bigl(\s^-_k(p),v\bigr),\cr
&c(p,f_1v)=\sum_kq^{1+2s(p)-2s_k(p)}c\bigl(\s^+_k(p),v\bigr).&(3.1.3)\cr
\endeq
These sums contain infinitely many terms.
Therefore, if we take an arbitrary element of the form (3.1.2),
the right-hand-side of (3.1.3) is not always well-defined.
We overcome this difficulty as follows:

Firstly, we make use of the fact
that the generators of $\uq$ come in pairs,
$(e_0,f_1)$ and $(e_1,f_0)$,
that generate identical paths in the sum (3.1.3).
The only difference is in the
factors of $q$ that they obtain.
But we will see that these factors of
$q$
are independent of
$k$
if
$k$
is large. Hence, though the action of a single generator
gives rise to an infinite number of terms,
the actions of certain linear combinations of
two generators, which we will denote by
$g_+$
and
$g_-$
in the below,
generate only a finite number of terms in the sum.

Secondly, if we restrict our attention to the highest weight vector
and the descendants, we can compute the actions of the generators
in closed form by using the highest weight condition (3.2.2) and the
usual commutation relations among the Chevalley generators.

\proclaim{Proposition 3.1}
Define
\eq
&g_+=f_0-t_1e_1,\ g_-=f_1-t_0e_0.&\cr
\endeq
The action of these operators on
$W$
is given by the following finite sum.
\eq
&c(p,g_+v)=(1-q^2)\sum_k[-1+2s(p)-2s_k(p)]c(\s^-_k(p),v),&(3.1.4)\cr
&c(p,g_-v)=(1-q^2)\sum_k[-2s(p)+2s_k(p)]c(\s^+_k(p),v),&(3.1.5)\cr
\endeq
where
$\displaystyle[n]={q^n-q^{-n}\over q - q^{-1}}$.

\noindent
This is immediate from (3.1.3). The finiteness follows from
the fact that the asymptotic values of
$s_k(p)$
in (3.1.4) and (3.1.5) when
$k\rightarrow\infty$
are
$s(p)-{1\over2}$
and
$s(p)$.

\subsec<3.2.|The highest weight vector>
By the highest weight vector we mean a linear combination of paths
\eq
&|\v\r=\sum_kc(p,\v)|p\r&(3.2.1)\cr
\endeq
that satisfies
\eq
&e_0|\v\r=0,\ e_1|\v\r=0.&(3.2.2)\cr
\endeq
Since we are aiming at the embedding
\eq
&V(\La_0)\rightarrow W,&\cr
\endeq
we also require
\eq
&t_0|\v\r=q|\v\r,\ t_1|\v\r=|\v\r,\ f_0^2|\v\r=0,\ f_1|\v\r=0.&(3.2.3)\cr
\endeq
We have, in particular, that
$c(p,\v)=0$
unless
$s(p)=0$.
For simplicity we abbreviate
$c(p,\v)$
to
$c(p)$
in this section.
In order to get
$c(p)$,
we may solve (3.2.2) perturbatively in
$q$, just as we did for (2.5.1).
However, using the observation we made above,
it is possible to compute
$c(p)$
exactly.

\sub<Example>
Let us compute $c(1)$.
We use
$(f_1-t_0e_0)|\v\r=0$.
Applying (3.1.5) for
$p=(---+-+\cdots)$,
we get
\eq
&[2]c(1)+c(\b)=0.\cr
\endeq
Hence $c(1)=-\e$.
\medskip
In general, we proceed as follows.
The equations (3.2.2-3) imply
\eq
&g_+^2|\v\r=0,\ g_-|\v\r=0.&(3.2.4)\cr
\endeq
Therefore, by using (3.1.4-5) we have
\eq
&\sum_k[2+2s_k(p)]c\bigl(\s^+_k(p)\bigr)=0
\hbox{ for $p$ such that $s(p)=-1$},&(3.2.5)\cr
&\sum_{k_1<k_2}[2-2s_{k_1}(p)][3-2s_{k_2}(p)]
c\bigl(\s^-_{k_1}\s^-_{k_2}(p)\bigr)=0
\hbox{ for $p$ such that $s(p)=2$.}&\cr
&&(3.2.6)\cr
\endeq

The question now is whether we can solve the above
equations simultaneously.
In the following, we show that the system of linear
equations (3.2.5-6) is
of block-triangular form.
We conjecture that the diagonal blocks are invertible.
Let
$\P_{l,s}$
be the set of paths
$p$
such that
$l(p)=l$
and
$s(p)=s$.
Then we have bijections
\eq
\P_{l,0}&\sim\P_{l,-1}\hbox{ if $l\ge2$ is even,}\cr
&\sim\P_{l,2}\hskip7pt\hbox{ if $l\ge3$ is odd.}\cr
\endeq
In fact, the cardinality is equal to
$\ \displaystyle {l-1\choose[{l\over2}]-1}$.
The bijection from
$\P_{l,-1}\ (l:\hbox{ even})$
or
$\P_{l,2}\ (l:\hbox{ odd})$
to
$\P_{l,0}$
is constructed as follows.
Take
$p\in\P_{l,-1}$
or
$p\in\P_{l,2}$.
Consider
$\bigl(-p(1),-p(2),\ldots,-p(l-1)\bigr)$,
and apply the procedure (2.4.1) to this finite sequence,
then we get a list
$a\in\P_{l,0}$.
For example, the correspondence up to
$l=4$
goes as
\eq
&(-\ -\ -\ +\ -\ +\cdots)\leftrightarrow(1),&\cr
&(+\ +\ +\ +\ -\ +\cdots)\leftrightarrow(01),&\cr
&(+\ -\ -\ -\ -\ +\cdots)\leftrightarrow(001),&\cr
&(-\ +\ -\ -\ -\ +\cdots)\leftrightarrow(11),&\cr
&(-\ -\ +\ -\ -\ +\cdots)\leftrightarrow(12).&\cr
\endeq
If
$l\bigl(\s^+_k(p)\bigr)>l(p)$ in (3.2.5),
then we have
$s_k(p)=-1$,
and if
$l\bigl(\s^-_{k_1}\s^-_{k_2}(p)\bigr)>l(p)$ in (3.2.6),
then we have
$s_{k_2}(p)={3\over2}$.
Therefore, the system of linear equations is of block-triangular form
with respect to the length of paths.

The coefficients of the first few equations are as follows.
\eq
\bordermatrix{
&\b&1&01&001&11&12&101&0001&011&012&\cr
--&1&[2]&0&0&0&0&0&0&0&0&\cr
+++&[2]&1&[2]^2&0&0&0&0&0&0&0&\cr
+---&0&1&0&0&[2]&[3]&0&0&0&0&\cr
-+--&1&0&0&[2]&0&[2]&0&0&0&0&\cr
--+-&0&0&1&1&[2]&0&0&0&0&0&\cr
-++++&[2]&0&[3]&1&0&0&0&[2]^2&[2][3]&[3]^2&\cr
+-+++&0&[2]&[2]&0&1&0&[2]^2&0&[2]^2&[2][3]&\cr
++-++&[2]&1&0&0&0&1&[2]&[2]^2&0&[2]^2&\cr
+++-+&0&0&0&[2]&1&0&1&[2]&[2]^2&0&\cr}
\endeq

The following is the list of
$c(p)$
for
$|\v\r$
up to
$l(p)=6$.

\settabs\+&\hbox{$|00001\r$\qquad}&\hbox{$\Xp$}\cr
\noindent$l=0$
\+&\hbox{$|\b\r$}&\hbox{$\Xa$}\cr
\noindent$l=2$
\+&\hbox{$|1\r$}&\hbox{$\Xb$}\cr
\noindent$l=3$
\+&\hbox{$|01\r$} &\hbox{$\Xc$}\cr
\noindent$l=4$
\+&\hbox{$|001\r$}&\hbox{$\Xd$}\cr
\+&\hbox{$|11\r$} &\hbox{$\Xe$}\cr
\+&\hbox{$|{12}\r$} &\hbox{$\Xf$}\cr
\noindent$l=5$
\+&\hbox{$|0001\r$}&\hbox{$\Xg$}\cr
\+&\hbox{$|011\r$}&\hbox{$\Xh$}\cr
\+&\hbox{$|0{12}\r$}&\hbox{$\Xi$}\cr
\+&\hbox{$|101\r$}&\hbox{$\Xj$}\cr
\noindent$l=6$
\+&\hbox{$|00001\r$}&\hbox{$\Xk$}\cr
\+&\hbox{$|0011\r$}&\hbox{$\Xl$}\cr
\+&\hbox{$|00{12}\r$}&\hbox{$\Xm$}\cr
\+&\hbox{$|0101\r$}&\hbox{$\Xn$}\cr
\+&\hbox{$|1001\r$}&\hbox{$\Xo$}\cr
\+&\hbox{$|111\r$}&\hbox{$\Xp$}\cr
\+&\hbox{$|112\r$}&\hbox{$\Xq$}\cr
\+&\hbox{$|{12}1\r$}&\hbox{$\Xr$}\cr
\+&\hbox{$|{12}2\r$}&\hbox{$\Xs$}\cr
\+&\hbox{$|{12}3\r$}&\hbox{$\Xt$}\cr

\subsec<3.3.|Descendants>
The descendants are by definition the vectors obtained from
the highest weight vector
$|\v\r$
by the actions of
$f_0$ and $f_1$.
We give a recursive formula for getting the descendants.
Note first that by (3.1.4-5) the action of
$g_+$
and
$g_-$
are finite on the space of vectors of the form (3.1.2).
In particular, the space of vectors created upon
$|\v\r$, which we denote by
$U$,
is well-defined. Let us give a recursive formula
for the actions on $U$, of the Chevalley generators of
$\uq$.
We use the notation
$\{x\}={\displaystyle {x-x^{-1}\over q-q^{-1}}}$.
Let
$|v\r\in U$
be a monomial of the form
\eq
&|v\r=g_{m_1}\cdots g_{m_k}|\v\r.&\cr
\endeq
If
$k\neq0$,
we set
$|v'\r=g_{m_2}\cdots g_{m_k}|\v\r$.
We have
\eq
t_0|v\r&=q^{1-2\sum_im_i}|v\r,\cr
t_1|v\r&=q^{2\sum_im_i}|v\r,\cr
e_0|v\r&=0\hskip1.6in\hbox{ if }k=0,\cr
&=\bigl(\{t_0\}+f_0e_0-e_0t_1e_1\bigr)|v'\r\quad\hbox{ if }m_1=+,\cr
&=\bigl(f_1e_0-e_0t_0e_0\bigr)|v'\r\hskip40pt\hbox{ if }m_1=-,\cr
\endeq
\eq
e_1|v\r&=0\hskip1.6in \hbox{ if }k=0,\cr
&=\bigl(f_0e_1-e_1t_1e_1\bigr)|v'\r\hskip40pt\hbox{ if }m_1=+,\cr
&=\bigl(\{t_1\}+f_1e_1-e_1t_0e_0\bigr)|v'\r\quad\hbox{ if }m_1=-,\cr
f_0|v\r&=(g_++t_1e_1)|v\r,\cr
f_1|v\r&=(g_-+t_0e_0)|v\r.&(3.3.1)\cr
\endeq

Let
$d$
be the derivation of
$\uq$
that satisfies
\eq
&[d,t_i]=0,\ [d,e_i]=-\delta_{i0}e_i,\ [d,f_i]=\delta_{i0}f_i.&(3.3.2)\cr
\endeq
We define the action of
$d$
on
$U$
by
$d|\v\r=0$.

Take
$s\in\Z$ and $\eg\in\Z_{>0}$.
An descendant of spin
$s$
and energy
$\eg$
is a vector
$|v\r$
of
$U$
satisfying
$t_1|v\r=q^{2s}|v\r,\ d|v\r=\eg|v\r$.
Let us denote the space of such vectors by
$U_{s\eg}$.

\sub<Example>
Let us calculate a few coefficients of
$f_0|\v\r$
and
$f_1f_0|\v\r$.
We have
\eq
&f_0|\v\r=g_+|\v\r.\cr
\endeq
{}From (3.1.4) a first few coefficients read as
\eq
c(++-+-+-+\cdots,f_0|\v\r)&=(1-q^2)c(\b)\cr
&=(1-q^2)\cdot1,\cr
c(+-++-+-+\cdots,f_0|\v\r)&=(1-q^2)\{c(01)+c(1)\}\cr
&=(1-q^2)(-2\e+\e^3),\cr
c(+-+-++-+\cdots,f_0|\v\r)&=(1-q^2)\{c(011)+c(101)+c(11)\}\cr
&=(1-q^2)(5\e^2-9\e^4+9\e^6-5\e^8+\e^{10}),\cr
c(-+++-+-+\cdots,f_0|\v\r)&=(1-q^2)\{[2]c(01)+c(\b)\}\cr
&=(1-q^2)\e^2.\cr
\endeq

Similarly, we have
\eq
&f_1f_0|\v\r=g_-g_+|\v\r+q|\v\r.\cr
\endeq
A first few coefficients are
\eq
&c(\b,f_1f_0|\v\r)=q,\cr
&c(1,f_1f_0|\v\r)=(1-q^2)^2-q\e,\cr
&c(01,f_1f_0|\v\r)=-(1-q^2)^2\e^2-q(\e-\e^3),\cr
&c(12,f_1f_0|\v\r)=\{2(1-q^2)^2-q\e\}(\e^2-\e^4).\cr
\endeq
\def\b{\hbox{\rm\O}}
\def\zer<#1>{\underbrace{0\cdots0}_{#1}}
\def\tw(#1){\underbrace{2\cdots2}_{k}}
\def\eg{\omega}
\def\v{\hbox{\rm vac}}
\def\E{\epsilon}
\def\s{\sigma}

\beginsection 4.
Identity between the eigenvectors of $\H$ and the $\uq$-weight vectors

\subsec<4.1.|Conjecture>
In \S 2 we developed a scheme for diagonalizing the CTM Hamiltonian
$\H_{re}$
acting on
$W$,
and introduced the space of eigenvectors
$H_{s\eg}$
of spin
$s$
and
energy
$\eg$.
In \S 3 we developed a scheme for realizing the level-1 vacuum
representation of
$\uq$
in
$W$,
and introduced the space of weight vectors
$U_{s\eg}$
of spin
$s$
and
energy
$\eg$.

We set forth

\proclaim{Conjecture 4.1}
The eigenvectors of $\H_{re}$ form the level-1 vacuum
representation
$V(\Lambda_0)$
of the quantum affine Lie algebra $\uq$,
i.e., we have
$H_{s\eg}=U_{s\eg}$.

\noindent
As a corollary we deduce that
\eq
&\H_{re}=d&(4.1.1)
\endeq
on
$U$. ($d$ is the derivation (3.3.2).)

In this section we report on our checks on this conjecture.
We used mostly the following data.

\pn
\sub<Data for $H_{s\eg}$>
\smallskip
Data 1:\quad
The coefficients
$c_l(p,\u)$,
for
$p=(\zer<k-1>1)$
and
$0\le l\le 5,\ 1\le k\le 41-l(l+1)$,
obtained by solving (2.5.3).

Data 2:\quad
The coefficients
$c_l\bigl(p_1,v(p_2)\bigr)$,
for
$p_1,p_2\in\{p\mid s(p)=0, \eg(p)\le 4\}$
and
$0\le l\le 2$,
obtained by solving (2.6.4). (There are 12 such paths.)

\pn
\sub<Data for $U_{s\eg}$>
\smallskip
Data 3:\quad
The coefficients
$c(p,\v)$,
for
$p$
such that
$l(p)\le9$,
obtained by solving (3.2.4-5). (There are 126 such paths.)

Data 4:\quad
The coefficients
$c\bigl(p_1,v(p_2)\bigr)$
for
$p_1,p_2\in\{p\mid s(p)=0, \eg(p)\le 3\}$,
obtained by the action (3.3.1). (There are seven such paths.)

\subsec<4.2.|Checks>

We made the following five different checks.

\sub<Check 1: Comparison of Data 1 and Data 3>
\medskip
This check is particularly important because
$c(\zer<k-1>1,\u)$
are the renormalization terms
$R_k$
in (2.5.4).
Note that Data 1 contain only first 6 nontrivial coefficients $(0\le k\le 5)$,
but of longer paths than Data 2, which are exact but limited to
the paths of length less than or equal to 9.
We checked the coincidence of these two sets of data
within the common region of accuracy and length.

Here are the relevant part of Data 3 in addition to those given in \S 3.

\def\Ya
{2\E^{19}-9\E^{17}+19\E^{15}-21\E^{13}+12\E^{11}-2\E^{9}-\E^{7}+\E^{3}-\E}
\def\Yb{-2\E^{25}+17\E^{23}-64\E^{21}+146\E^{19}-228\E^{17}+247\E^{15}}
\def\Yc{-179\E^{13}+79\E^{11}-16\E^{9}-\E^{7}+\E^{3}-\E}
\def\Yd
{6\E^{33}-47\E^{31}+188\E^{29}-499\E^{27}+968\E^{25}-1433\E^{23}+1638\E^{21}}
\def\Ye
{-1420\E^{19}+893\E^{17}-375\E^{15}+86\E^{13}-2\E^{11}-2\E^{9}-\E^{7}
+\E^{3}-\E}
\medskip
\settabs\+&\hbox{$|00000001\r$\qquad}&\hbox{$\Yd$}\cr
\+&\hbox{$|000001\r$}&\hbox{$\Ya$}\cr
\smallskip
\+&\hbox{$|0000001\r$}&\hbox{$\Yb$}\cr
\+&&\hbox{$\Yc$}\cr
\smallskip
\+&\hbox{$|00000001\r$}&\hbox{$\Yd$}\cr
\+&&\hbox{$\Ye$}\cr
\medskip

Data 1 and Data 3 also show a strong evidence that
$c_l(\zer<k>1,\u)$
is independent of
$k$
for
$k\ge\max(2l-1,0)$.
In general, we conjecture that the following two asymptotic properties.
We use the list notation (2.4.1) for the spin-0 paths.
\medskip
\item{(A)}
For any path
$p$,
the coefficient
$c_l(\zer<k>p,\u)$
is independent of
$k$
for
$k\ge\max(2l-1,0)$.
\item{(B)}
For any paths
$p$ and $p'$,
we have
$c_l(p\zer<k>p',\u)
=\sum_jc_j(p,\u)c_{l-j}(\zer<k+l(p)>p',\u)$
for
$k\ge2l+1$.
The case when
$p'=(1)$ was already mentioned in \S 2.

\sub<Check 2: $\H|\v\r=0$ by using Data 3>

Let us abbreviate
$c(p,\v)$
by
$c(p)$.
Given a spin-0 path, we wish to check
\eq
&\bigl\{\eg(p)-\e\sum_{k\ge1}kc(\zer<k-1>1)\bigr\}c(p)
+\e\sum_{k\ge1}kc(Q_kp)=0.\cr
\endeq

Let us discuss a simple example.

\sub<Example>
Let us consider (4.2.1) for
$p=(+\ -\ -\ +\ -\ +\ \cdots)$.
We have
$\eg(1)=1$,
and
$c(\zer<k-1>1)=-\e,-\e+\e^3,-\e+\e^3-\e^5,\ldots$
for
$k=1,2,3,\dots$,
as given previously.
The second sum in (4.2.1) reads as
$c(\b)+3c(11)+4c(101)+5c(1001)+\cdots$.
The relevant terms which are not included in the table in \S 3 are

\def\Za{-3\E^{18}+14\E^{16}-34\E^{14}+48\E^{12}-40\E^{10}+18\E^{8}-3\E^{6}-\E^{4}+\E^{2}}
\def\Zb{2\E^{26}-17\E^{24}+64\E^{22}-148\E^{20}+240\E^{18}-280\E^{16}}
\def\Zc{+234\E^{14}-139\E^{12}+58\E^{10}-16\E^{8}+3\E^{6}-\E^{4}+\E^{2}}
\def\Zd{-8\E^{32}+60\E^{30}-242\E^{28}+682\E^{26}-1450\E^{24}+2387\E^{22}-3073\E^{20}}
\def\Ze{+3082\E^{18}-2365\E^{16}+1341\E^{14}-528\E^{12}+127\E^{10}-13\E^{8}-\E^{4}+\E^{2}}

\medskip
\settabs\+&\hbox{$|1000001\r$\qquad}&\hbox{$\Zd$}\cr
\+&\hbox{$|10001\r$}&\hbox{$\Za$}\cr
\smallskip
\+&\hbox{$|100001\r$}&\hbox{$\Zb$}\cr
\+&&\hbox{$\Zc$}\cr
\smallskip
\+&\hbox{$|1000001\r$}&\hbox{$\Zd$}\cr
\+&&\hbox{$\Ze$}\cr
\medskip

Since we can take account of only finitely many terms,
we must estimate the accuracy we can expect.
We use the paths of length less than or equal to 9.
Admitting the asymptotic factorization (B) mentioned above,
we can expect the cancellation
$\bmod\  \e^9$. This is in fact true.

Following a similar procedure we have checked that
$\H_{re}|\v\r=0$ within the accuracy expected from Data 3,
and also that the state
$f_0|\v\r$
is the first excited state
of $\H_{re}$ with eigenvalue 1. However, because the coefficients
are much more complicated for this state, we could perform
the computations only for all paths up to
length 8, and to lowest non-trivial order in $\e$.

\sub<Check 3: Comparison of Data 2 and Data 4>
\medskip
The consistency of the equation (2.6.4) is not obvious,
especially when the eigenvalue has non-trivial multiplicity.
Therefore, we computed Data 2 rather extensively,
i.e., the energy up to 4,
as a check of (2.6.4) itself. The result was in favor of our
consistency assumption.

Then we checked the coincidence of Data 2 and Data 4 within
the common region of accuracy and energy.
We omit the details of this check, since the most of the
data were created only in the memory of our computer.

\sub<Check 4: $e_0|\v\r=e_1|\v\r=0$ by using Data 3>
\medskip
Recall that the reason why we could evaluate coefficients
$c(p,\v)$
exactly, in \S 3,
was that we made use of (3.2.4) instead of (3.2.2-3).
In Check 4 that follows, we have to act with single generators
$e_0$ and $e_1$,
so we end up with infinite series (3.1.3), and must turn to perturbation
theory.

The equations read as
\eq
&\sum_kq^{2s_k(p)}c\bigl(\s^+_k(p),v\bigr)=0\hbox{ for }s(p)=-1,&(4.2.2)\cr
&\sum_kq^{-2s_k(p)}c\bigl(\s^-_k(p),v\bigr)=0\hbox{ for }s(p)=1.&(4.2.3)\cr
\endeq

Let us discuss a simple example.

\sub<Example>
Let us cosider (4.2.3) for
$p=(+\ +\ -\ +\ -\ +\ \cdots)$.
We are to check
\eq
&c(1)+c(12)+c(122)+\cdots=-q.&(4.2.4)\cr
\endeq

\pn
The datum available in addition to those given in \S 3 is
\smallskip
\noindent
$|1222\r$
\smallskip
$6\E^{25}-39\E^{23}+128\E^{21}-277\E^{19}+429\E^{17}
-485\E^{15}+396\E^{13}-225\E^{11}+81\E^{9}-14\E^{7}$
\smallskip
\noindent
Assuming that
\eq
&c(1\tw(k))=0\ \bmod\e^{2k+1},&\cr
\endeq
we can expect
\eq
&c(1)+c(12)+\cdots+c(1222)=-\e-\e^3-2\e^5-5\e^7\bmod\e^9,\cr
\endeq
which is certainly true.
\smallskip
In general, we have checked (4.2.2-3) up to the accuracy that
are expected from the data for the paths of length less than or equal to 9.

\sub<Check 5: (4.2.4) by using some additional data>
\medskip
Since we explicitly know the coefficients of only a finite
number of paths: those up to length 9, the accuracy of the
cancellations is rather limited. To make our check sure we
computed a few more terms in (4.2.4) by solving (2.5.3):
\eq
c(12222)&=-84\e^9+770\e^{11}-3529\e^{13}+\cdots,\cr
c(122222)&=-594\e^{11}+7722\e^{13}+\cdots,\cr
c(1222222)&=-4719\e^{13}+\cdots.\cr
\endeq
By using this we checked (4.2.4)
$\bmod\ \e^{15}$.

\subsec<4.3.|The commutation relations>

Since we are interested in the relation between $\H_{re}$,
and $\uq$, we wish to evaluate their commutators.
We have
\eq
&[\H_{re},e_i]=-\delta_{i0}e_i,\ [\H_{re},f_i]=\delta_{i0}f_i.&(4.3.1)\cr
\endeq

Each commutator maps a given path to an infinite superposition of paths.
We will show that in each case of (4.3.1), the commutator in the left-hand-side
gives the same superposition and the Chevalley generator in the
right-hand-side, for each choice of a path.

We neglect all scalar terms in $\H_{re}$, since their contributions
trivially cancel out in the commutators. We also discard the term
proportional to
$S$,
since its commutators with
$e_i$, $f_i$ are known.
So, we set
$\H'=P'+\e Q$,
and compute its commutators with
$e_i$, $f_i$.

{}From \S 2, we recall that $P' = \sum_{k = 1}^\infty k P'_k$,
and $Q = \sum_{k = 1}^{\infty} k Q_k$, where $P'_k$ and $Q_k$
are bi-local operators: they act on pairs of neighbouring spins (2.2.1-2).
{}From \S 3, we recall the action
of the co-multiplication of $e_i$ and $f_i$ on a path (3.1.1).
The latter is the sum of an infinite number of semi-infinite tensor
products. Let us use $e_{ik}$, $f_{ik}$
to denote single semi-infinite tensor products, with a generator
in the $k$-th position, e.g.,
\eq
&e_{1k} =
\underbrace{t_i^{-1}\o\dots\o t_i^{-1}}_{k-1}\o e_1\o1\o\dots\cr
\endeq

We will derive $[\H_{re},e_1]$ in some detail as an example.
Consider
$[\H'_k,e_{1k'}]$
where
$\H'_k=P'_k+\e Q_k$.
It vanishes unless $k'=k,k+1$. Therefore, we have two cases
to consider: $k'=k$ and $k'=k+1$.  We start with
$k'=k$. We compute the action of the commutator on the 4
possible spin-configurations
$(s\ s')\ (s,s'=\pm)$
on the $k$-th and the $(k+1)$-the
components. We start with
$(-\ -)$:
\eq
(P'_k+\e Q_k)e_{1k}(-\ -)&=(P'_k+\e Q_k)q^{-2s_k(p)}(+\ -)\cr
& =q^{-2s_k(p)}(+\ -)+\e q^{-2s_k(p)}(-\ +),\cr
-e_{1k}(P'_k+\e Q_k)(-\ -)&=-q^{-2s_k(p)}(+\ -).\cr
\endeq
Therefore we have,
$[\H'_k,e_{1k}](-\ -)=\e q^{-2s_k(p)}(-\ +)$.

Similarly, we obtain the following as a whole:
\eq
&[\H'_k,e_{1k}](-\ -)=\e q^{-2s_k(p)}(-\ +),&(4.3.2)\cr
&[\H'_k,e_{1k}](-\ +)=q^{-2s_k(p)}(+\ +),&(4.3.3)\cr
&[\H'_k,e_{1k}](+\ -)=-\e q^{-2s_k(p)}(+\ +),&(4.3.4)\cr
&[\H'_k,e_{1k}](+\ +)=0.&(4.3.5)\cr
\endeq
Repeating the same excercise for $e_{1\,k+1}$, we obtain:
\eq
&[\H'_k,e_{1\,k+1}](-\ -)=\e q^{-2s_{k}+1}(+\ -)-q^{-2s_{k}+1}(-\ +),
&(4.3.6)\cr
&[\H'_k,e_{1\,k+1}](-\ +)=-\e q^{-2s_{k}-1}(+\ +),&(4.3.7)\cr
&[\H'_k,e_{1\,k+1}](+\ -)=0,&(4.3.8)\cr
&[\H'_k,e_{1\,k+1}](+\ +)=0.&(4.3.9)\cr
\endeq
Note that in (4.3.2-9) the bracket changes a
$(-)$
into a
$(+)$
(if the result is not just zero.)

Next, we will consider the action of the commutator $[\H'_k,e_{1k'}]$
on an arbitrary but fixed path, say
$p$,
and collect the contributions given by (4.3.2-9) such that
the change of the
$(-)$
into the
$(+)$
takes place at the
$k$-th component.

Let us start with the case where the there spins
at the $(k-1)$-th, the $k$-th and the $(k+1)$-th components are
$(+\ -\ +)$.
The contributions to
$(+\ +\ +)$
are from (4.3.3), (4.3.4) and (4.3.7),
i.e.,
$-(k-1)\e q^{-2s_{k-1}(p)}$,
$kq^{-2s_k(p)}$ ,
$-k\e q^{-2s_k(p)-1}$,
respectively.
Summing up these three terms, and using
$s_k(p)=s_{k-1}(p)+{1\over2}$,
we get
$\e q^{-2s_{k-1}(p)}$.

This is exactly the coefficient that we would obtain under the action of
$\e qe_{1k}$.
This can be verified for the rest of the possible cases, and we deduce that
\eq
&[\H',e_1]=\e qe_1.\cr
\endeq

We have to pause at this point to remark that the individual actions
of the commutator give weights that contain the position
$k$ as a factor. All such dependences neatly cancel out, once we
sum the various contributions. When one performs the calculation,
such cancellations are totally mysterious.

But, they are precisely these cancellations that allow the final result
to coincide with that obtained from the action of the Chevalley generator, and
that contains no such multiplicative $k$-dependence.

The rest of the commutation relations can be proven in exactly the same way.
We obtain:
\eq
&[\H',e_0]=-\e q^{-1}e_0,\cr
&[\H',f_0]=\e q^{-1}f_0,\cr
&[\H',f_1]=-\e qf_1.\cr
\endeq

{}From these commutation relations, we can see that $\H_{re}$ acts
as a derivation of $\uq$.

As discussed above, these simple
commutators is at this point unexpected and mysterious.
Together with the direct checks that we discussed before,
they encouraged us to present Conjecture 4.1.

\bigskip
{\it Note added in proof.}\quad
After finishing this work, [FT] was brought to our attention,
where a study of CTM eigenvectors of the six-vertex model, both in the
ferro-electric $\Delta>1$, and anti-ferro-electric $\Delta<-1$
regimes, is presented.

The connection between our eigenstates, and those obtained in [FT]
remains to be understood. The point is that Frahm and Thacker
are considering the CTM eigenstates on the whole line, while our CTM
eigenstates are on the half line.

We also add [PS] in our reference, in which the commutation relations
between the row-to-row transfer matrix Hamiltonian and
$U_q\bigl(\goth{sl}\,(2)\bigr)$ is calculated on a finite lattice.
Note that our commutation relations are between the CTM Hamiltonian and
$U_q\bigl(\widehat{\goth{sl}}\,(2)\bigr)$.

\bigskip
\def\refu{\itemitem{[PS]}
V. Pasquier and H. Saleur,
Common structures between finite systems and conformal
field theories through quantum groups,
Nucl. Phys. B {\bf 330} (1990),  523-556.}
\def\refv{\itemitem{[FT]}
H. Frahm and H.B. Thacker,
Corner transfer matrix eigenstates for the six-vertex model,
preprint, University of Virginia, 1991.}
\def\reft{\itemitem{[SW]}
K. Sogo and M. Wadati,
Boost operator and its application to quantum Gelfand-Levitan
equation for Heisenberg-Ising chain with spin one-half,
Prog. Theor. Phys. {\bf 69} (1983), 431.}
\def\refb{\itemitem{[B1]}
R.J. Baxter,
Exactly Solved Models in Statistical Mechanics,
1982 Academic.
\vskip 1.5mm}
\def\refa{\itemitem{[ABF]}
G.E. Andrews, R.J. Baxter and P.J. Forrester,
Eight-vertex SOS models and generalized Rogers-Ramanujan identities,
J. Stat. Phys. {\bf 35} (1984), 193-266.
\vskip 1.5mm}
\def\reff{\itemitem{[DJKMO1]}
E. Date, M. Jimbo, A. Kuniba, T. Miwa and M. Okado,
Exactly solvable SOS models:
Local height probabilities and theta function identities,
Nucl, Phys. B{bf 290} [FS20] (1987), 231-273,
II. Proof of the star-triangle relation and combinatorial identities,
Adv. Stud. Pure Math. {\bf 16} (1988), 17-122.
\vskip 1.5mm}
\def\refg{\itemitem{[DJKMO2]}
Date, E., Jimbo, M., Kuniba, A., Miwa, T. and Okado, M,
One dimensional configuration sums in vertex models and affine Lie
algebra characters,
Lett. Math. Phys., {\bf 17} (1989), 69-77.
\vskip 1.5mm}
\def\refo{\itemitem{[(KMN)${}^2$]}
S.J. Kang, M. Kashiwara, K.C. Misra, T. Miwa, T. Nakashima and A. Nakayashiki,
Affine crystals and vertex models, preprint, RIMS-828 (1991).
\vskip 1.5mm}
\def\refp{\item{[MM]} K.C. Misra and T. Miwa,
Crystal base for the basic representation
of $U_{q}(\hat sl(n))$,
Commun. Math. Phys. {\bf 134} (1990), 79-88.
\vskip 1.5mm}
\def\refm{\itemitem{[JMMO]} M. Jimbo, K.C. Misra, T. Miwa and M. Okado,
Combinatorics of representations of $U_{q}(\hat sl(n))$ at $q=0$,
Comm. Math. Phys. {\bf 136} (1991), 543-566.
\vskip 1.5mm}
\def\refh{\itemitem{[Dr]} V.G. Drinfeld,
Quantum groups, Proc. ICM Berkeley (1987), 798-820.
\vskip 1.5mm}
\def\refl{\itemitem{[J]} M. Jimbo,
Introduction to the Yang-Baxter equation,
Inter. J. Mod. Phys. A, {\bf 4} (15) (1989), 3759-3777.
\vskip 1.5mm}
\def\refc{\itemitem{[B2]}
R.J. Baxter,
Corner Transfer Matrices of the Eight-Vertex Model,
J. Stat. Phys. {\bf 15} (1976), 485-503; {\bf 17} (1977), 1-14.
\vskip 1.5mm}
\def\refd{\itemitem{[Da1]}
B. Davies,
Corner transfer matrices for the Ising model,
Physica A {\bf 154} (1988), 1-20.
\vskip 1.5mm}
\def\refe{\itemitem{[Da2]}
B. Davies,
On the spectrum of six-vertex corner transfer matrices,
Physica A {\bf 159} (1989), 171-187.
\vskip 1.5mm}
\def\refn{\itemitem{[K]}
M. Kashiwara,
Crystalizing the $q$-analogue of universal enveloping algebras,
Commun. Math. Phys. {\bf 133} (1990), 249-260;
On crystal bases of the $q$-analogue of universal enveloping algebras,
Duke Math. J. {\bf 63} (1991), 465-516.
\vskip 1.5mm}
\def\refs{\itemitem{[TP]}
T.T. Truong and I. Peschel,
Diagonalization of finite-size corner transfer matrices and related
spin chains, Z. Phys. B {\bf 75} (1989), 119-125.
\vskip 1.5mm}
\def\refk{\itemitem{[IT]}
H. Itoyama and H.B. Thacker,
Integrability and Virasoro symmetry of the noncritical
Baxter/Ising model,
preprint, Fermilab, 1988.
\vskip 1.5mm}
\def\refw{\itemitem{[Te]}
M.G. Tetel'man,
Lorentz group for two-dimensional integrable lattice systems,
Zh. Eksp. Teor. Fiz. {\bf 82} (1982), 528-535.}
\def\refr{\itemitem{[T]}
H.B. Thacker,
Corner transfer matrices and Lorentz invariance on a lattice,
Physica D {\bf 18} (1986), 348-359.
\vskip 1.5mm}
\def\refq{\itemitem{[SB]}
H. Saleur and M. Bauer,
On some relations between local height probabilities and conformal invariance,
Nucl. Phys. B {\bf 320} (1989), 591-624.
\vskip 1.5mm}
\def\refi{\itemitem{[F]}
O. Foda,
A relation between off-critical local height probabilities,
and critical pertition functions in a finite geometry,
preprint, Univ. of Nijmegen, 1991.
\vskip 1.5mm}
\def\refz{\itemitem{[DP]}
B. Davies and P. A. Pearce,
Conformal invariance and critical spectrum of corner transfer matrices,
J. Phys. A {\bf 23} (1990), 1295-1312.
\vskip 1.5mm}

{\bf References}
\bigskip
{\rm
\refa
\refb
\refc
\refd
\refe
\refh
\reff
\refg
\refz
\refi
\refk
\refl
\refm
\refn
\refo
\refp
\refq
\reft
\refw
\refr
\refs
\refv
\refu
}
%
%
%
%
\end